\documentclass[twocolumn,showpacs,preprintnumbers,amsmath,amssymb]{revtex4}
\usepackage{graphicx}
\usepackage{dcolumn}
\usepackage{bm}

\begin{document}

\bibliographystyle{alpha}
\title{ Morphological Instability Induced by the Interaction
of a Particle with a Solidifying Interface}
\author{ Layachi  Hadji}
\affiliation{The University of Alabama,
Department of Mathematics,
Tuscaloosa, Alabama 35487}
 
\date{\today}

\begin{abstract}
We show that the interaction of a particle with a 
directionally solidified interface induces the onset of morphological
instability provided that the particle-interface distance falls below a 
critical value. This instability occurs at pulling velocities that are
below the threshold for the onset of the Mullins-Sekerka instability.
The expression for the critical distance reveals that
this instability is manifested only for certain combinations of the physical
and processing parameters. Its occurence is attributed to the reversal of the
thermal gradient in the melt ahead of the interface and behind the particle.
\end{abstract}
\widetext

\maketitle

The freezing of a liquid with a dispersed phase takes place in numerous
natural and industrial processes.  Some examples include
the formation of ice lenses that result from the freezing of soil water
\cite {Jac58}, the freezing of biological cell suspensions in a 
cryopreservation experiment \cite {Ish94}, the decontamination of metallic
pollutants from soils \cite {Gay02}, the growth of Y123 superconductors by
the undercooling method \cite{End96} and the manufacture of particulate
reinforced metal matrix composites (PMMC) \cite {Cly91}. The properties of
these composite materials are enhanced by the addition of the dispersed
elements.The freezing of a liquid suspension is associated with the interaction
of the constituents of the dispersed phase with a solidifying interface. 
The first systematic
study of this interaction was carried out by Uhlmann {\it et al.} \cite{Uhl64}. 
They demonstrated the existence of a critical value for the growth rate below
which the inclusions are pushed by the moving interface, and above which they
are engulfed by the interface and incorporated into the solid. Consequently,
very low growth rates are conducive to particles being pushed by the interface,
while high growth rates are conducive to particle engulfment. \\

The presence of an inclusion  in the melt near a solid-liquid interface
introduces locally a change, albeit small,  in the thermal gradient ahead of
the solid front. This, in turn, introduces a small deformation in the profile
of the interface. The difference in the thermal conductivities of
the melt and particle stands out as the cause for this interfacial 
deflection \cite{Zub73,Aac77,Sen97,Had01}. Imagine a situation wherein a
solid is growing antiparallel to the direction of the heat flux and toward an 
inclusion  that is less heat conducting than the melt in which it is
immersed. Then, as the width of the gap separating the inclusion from 
the solid front decreases, heat becomes more easily evacuated from the solid
phase than from ahead, leading to a local reversal of the thermal gradient
and, consequently, to the local destabilization of the interface.
As long as the particle remains pushed, the disturbance grows and propagates
radially with decreasing magnitude.\\

\begin{figure}
\includegraphics[height=2.00in]{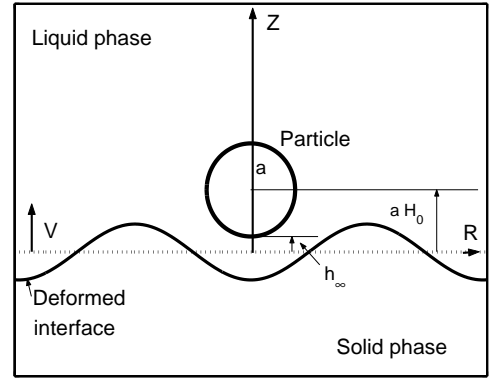}
\caption{\label{fig:epstart}
Sketch of a particle of radius $a$ immersed in a melt near a deformable
solid-liquid interface; $a H_0$ is the particle-interface
distance measured from the particle's center to the planar solid front,
$V$ is the interface growth rate, $h_{\infty}$ is the gap width that separates
the lowest point on the particle from the planar interface, and $R$ represents
the radial coordinate taken along the planar interface.}
\label{f.1}
\end{figure}

This hand-waving argument is made more precise in this paper.
The situation considered here is illustrated schematically in Fig. 1.
An axisymmetric spherical particle of radius $a$ is submerged in a binary
alloy in such a way that the distance from its center  to 
the planar solid-liquid interface that emerged from the
directional solidification of a dilute binary mixture is $ a H_0$. We assume 
a microgravity environment.  The problem
is described by the diffusion of solute equation in the melt and the heat 
conduction equations in the melt, solid phase and particle  in a frame that
is moving with velocity $V$ in the upward $Z-$direction; $Z$ is the vertical 
coordinate ($Z>0$ in the liquid region)
and $R$ is the radial coordinate taken along the planar solid-liquid
interface. The diffusion of solute in the solid phase is neglected.
The boundary conditions at the solid-liquid interface account for: (i) 
curvature and solute undercoolings in the equation describing the equilibrium
temperature, (ii) the heat balance, and (iii) the conservation of mass.
At the particle-melt boundary, $[Z-(a+h_{\infty})]^2+R^2=a^2$, we
impose the zero mass flux and the continuity of the temperature and heat flux;
$h_{\infty}$ denotes the width of the gap between the particle and the planar
interface. Far away from the solid front, the imposed thermal gradients in the
solid and liquid phases are $G^{(L)}$ and $G^{(S)}$, respectively, while the 
concentration of solute is maintained at $C_{\infty}$. 
The nondimensionalisation that we adopt makes use of the particle's
radius $a$  as length scale \cite{not01}, $a/V$, the melting point $T_m$,
the growth rate $V$, and the concentration far away from the front, 
$C_{\infty}$, as scales for time, temperature, velocity and concentration,
respectively. We have
\begin{equation}
\epsilon ( \frac {\partial C}{\partial t} -\frac {\partial C}{\partial z})
 = {\Delta}_{r} C + \frac {\partial^2 C}{\partial z^2},
\label{eq:one}
\end{equation}
\begin{equation}
\epsilon \lambda^{(q)} ( \frac {\partial T^{(q)}}{\partial t} -
\frac {\partial T^{(q)}}{\partial z} ) =
{\Delta}_{r} T^{(q)} + \frac {\partial^2 T^{(q)}}{\partial z^2},
\label{eq:two}
\end{equation}
where $C$ is the solute concentration, $T$ is the temperature, the 
superscript $(q)$ stands symbolically for particle $(q = P)$, solid $(q=S)$
and liquid $(q=L)$, $\lambda^{(q)} = D/D^{(q)}$, where $D$ and $D^{(q)}$ are
the coefficients of solute and thermal diffusion, respectively,
$\epsilon = aV/D$ is the Peclet number, and $\Delta_r$ is the Bessel   
differential operator, $(1/r)\partial/{\partial r} (r \partial /\partial r)$.
The corresponding boundary conditions
are as follows: at the solid-liquid interface,
\begin{equation}
T^{(S)} = T^{(L)} = 1 - \sigma \kappa + M C,
\label{eq:three}
\end{equation}
\begin{equation}
{\cal S} v_{n} = (k \nabla T^{(S)} - \nabla T^{(L)}) \cdot {\bf n}_i,
\label{eq:four}
\end{equation}
\begin{equation}
\epsilon C (1 - K)v_n = - \nabla C \cdot {\bf n}_i.
\label{eq:five}
\end{equation}
At the particle's surface, $(z-H_0)^2+r^2=1$, the continuity of temperature
and of the heat flux imply,
\begin{equation}
T^{(L)} = T^{(P)}, \quad
(\alpha \nabla T^{(P)} - \nabla T^{(L)}) \cdot {\bf n}_P =0,
\label{eq:six}
\end{equation}
while the zero mass flux condition yields $\nabla C \cdot {\bf n}_P =0$.
The far field conditions are 
$ {\partial T^{(L)}}/{\partial z} \rightarrow  G_L $ and $C=1$ as 
$ z \rightarrow \infty$, and $ {\partial T^{(S)}}/{\partial z} 
\rightarrow  G_S$ as $ z \rightarrow -\infty$.
The symbols that appear in the above equations are defined as follows:
$\sigma = \sigma_{SL}/a L$ is the surface energy parameter, where 
$\sigma_{SL}$ is the interface excess free energy and $L$ is the latent
heat of fusion per unit volume;  $M = m C_{\infty}/T_m$, is the morphological
 parameter, where $m$ is the
liquidus slope; ${\cal S} = L V a/(T_m k^{(L)})$ is the Stefan number;
$k = k^{(S)}/k^{(L)}$; $\alpha = k^{(P)}/k^{(L)}$, where $k^{(q)}$ is the
coefficient of thermal conductance; $K$ is the segregation coefficient;
$v_n$ is the normal growth velocity; ${\bf n}_i$ and ${\bf n}_P$ are the
unit normal vectors pointing into the melt at the interface and particle's
surface, respectively; $G_L = a G^{(L)}/T_m$ and $G_S = a G^{(S)}/T_m$, where
$G^{(q)}$ is the imposed (dimensional)
thermal gradient, and $\kappa$ is the curvature
taken to be positive when the center of curvature lies in the soild phase.\\

The above set of equations ( with $\epsilon =0$) \cite{not02}
admits a base state with a planar interface 
growing at constant speed.
The determination of the thermal fields in an infinite medium in which a
spherical inclusion has been embedded has been carried out \cite{Car59}. Here,
we make use of the method of images \cite{Che98} 
to calculate the thermal fields in the
melt and particle that satisfy the equilibrium temperature condition at the
planar interface, Eq. (3), and the conditions at the particle's surface,
Eq. (6). Then an expression for the thermal field in the solid is derived.
This expression must satisfy the temperature condition at the planar interface,
Eq. (3), and must also support a planar interface growing with constant
speed, Eq. (4). We have
\begin{equation}
C_B(r,z)=1, \quad {\mbox{in the melt}},
\end{equation}
\begin{equation}
T_B^{(L)}(r,z) = 1 + G_L z + M + U(r,z),
\quad {\mbox{in the melt}},
\label{eq:nine}
\end{equation}
\begin{equation}
T_B^{(S)}(r,z) = 1 + G_S z + M + {U(r,z) \over k},
\quad {\mbox{in the solid}},
\label{eq:ten}
\end{equation}
\begin{figure}
\includegraphics[height=1.8in]{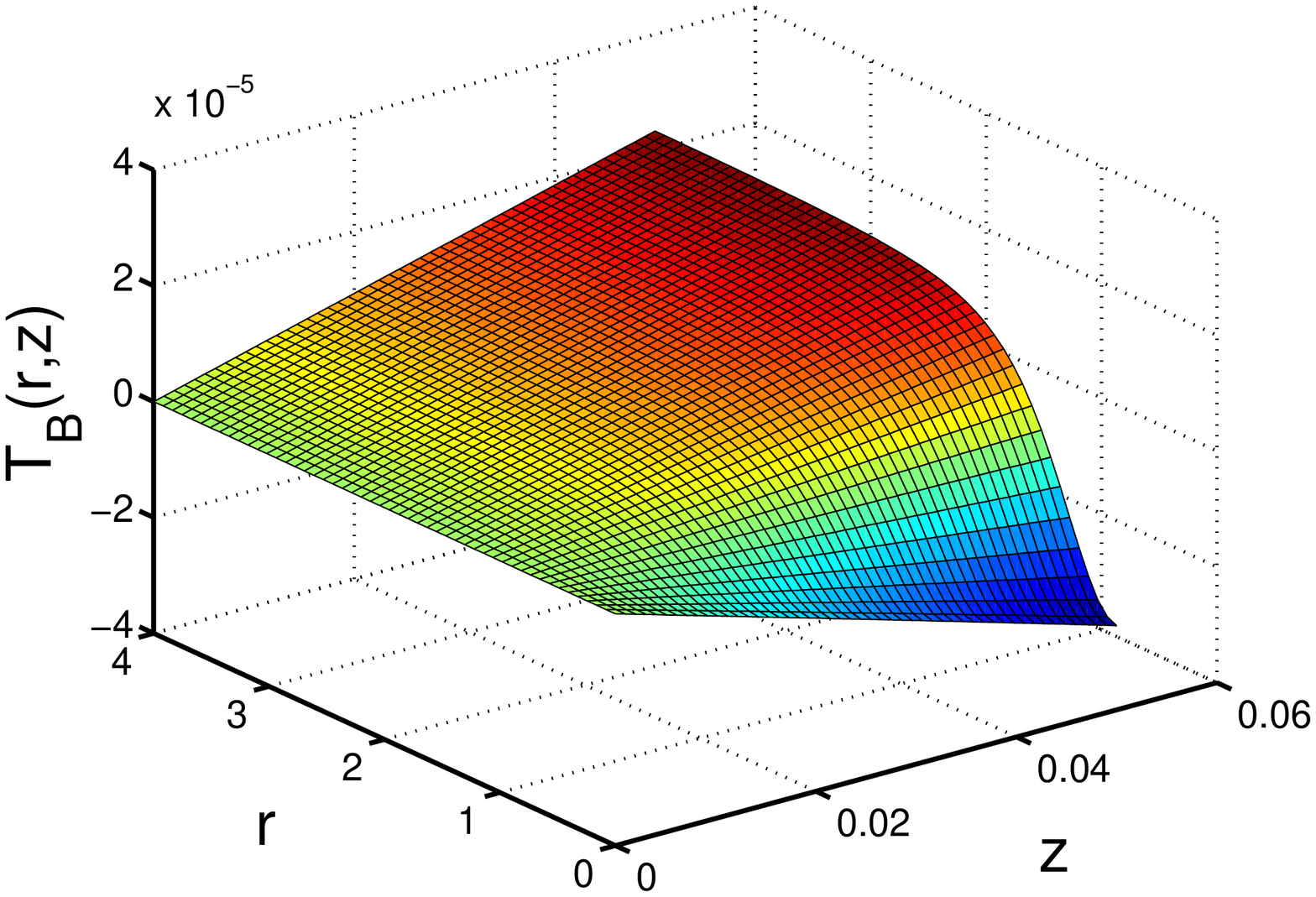}\\
\includegraphics[height=1.8in]{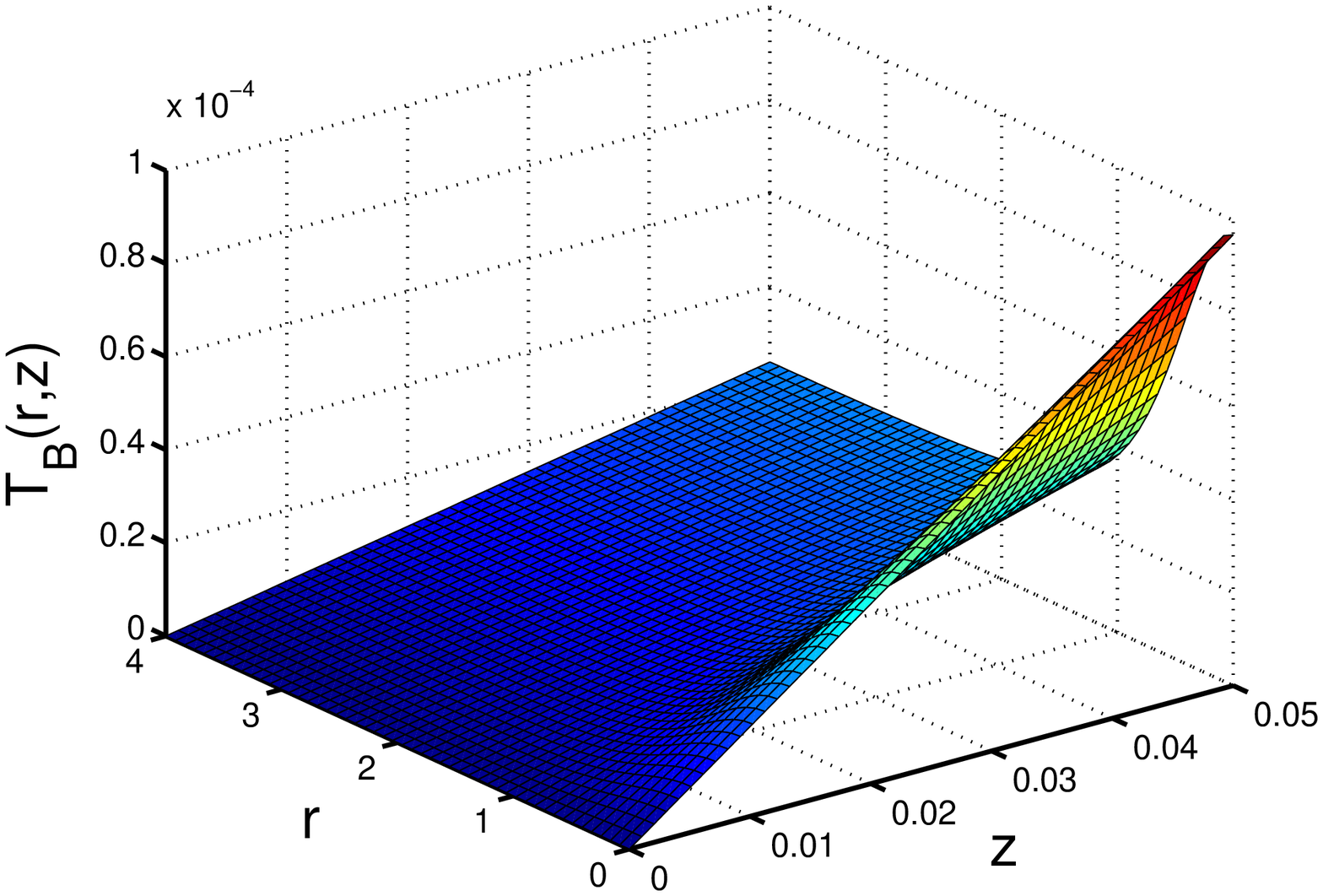}
\caption{\label{fig:epstart}
 Plot of the basic temperature profile in the melt (Eq. (8)) 
for a particle of radius $a= 10 \mu m$, dimensionless gap thickness
$h_{\infty}/a = 0.05$ and thermal conductance ratios
$\alpha=0$ (left) and $\alpha=10$ (right). Note the reversal
in the temperature gradient in the gap separating the particle from the
interface for the insulating particle case. The profile is evaluated
using parameter values for a succinonitrile-acetone system (SCN-ACE)
\cite {Kur92,Ste95}.}
\end{figure}
\begin{equation}
T_B^{(P)}(r,z) = {3  G_L z \over 2+\alpha} + 1 +M, \quad
{\mbox{in the particle}}.
\label{eq:eleven}
\end{equation}
where
\begin{equation}
U(r,z) = G_{L}({ 1-\alpha \over 2+\alpha } )
          \sum_{\scriptstyle n=-\infty\atop
          \scriptstyle n \ne 0}^{\infty} 
                             {z + n H_0 \over [(z + nH_0)^2+r^2]^{3/2}}.
\label{eq:twelve}
\end{equation}
The plot of the basic profile for the melt's temperature, Eq. (8),
is shown in Fig. 2. It pertains to an axisymmetric particle  of 
radius $a=10 \mu m$ whose center is on the $z-$axis and whose separation 
from the interface is $0.5 \mu m$. It depicts
an unstable configuration, wherein the thermal gradient in the gap is reversed
due to the thermally insulating character of the particle. The configuration
is, however, stable for the case of a highly conducting particle.\\

In order to examine the stability of the base state, we first superimpose
axisymmetric, time-dependent infinitesimal disturbances $\theta^{(q)}$, $c$
and $\eta$ upon the basic state solutions $T_B^{(q)}$, $C_B$, and the planar  
interface, respectively. We let
\begin{equation}
[T^{(q)},C] = [T_B^{(q)}, C_B] + \epsilon [\theta^{(q)}, c],
\end{equation}
in Eqs. (1)-(6). The resulting equations are then linearized with respect to
the disturbances. The following perturbed problem is obtained,
\begin{equation}
\Delta_r \theta^{(q)} + \frac {\partial^2 \theta^{(q)}}{\partial z^2}
= - \lambda^{(q)} \frac {\partial T_B^{(q)}}{\partial z},
\label{eq:twelveb}
\end{equation}
\begin{equation}
\Delta_r c + \frac {\partial^2 c}{\partial z^2} = 0,
\label{eq:thirteen}
\end{equation}
\begin{equation}
\theta^{(L)} = - G_L \eta + \sigma \Delta_r \eta - 
F(r) \eta + M c, \quad z=0,
\label{eq:fourteen}
\end{equation}
\begin{equation}
\theta^{(S)} = - G_S \eta + \sigma \Delta_r \eta - 
\frac {F(r)}{k} \eta + M c, \quad z=0,
\label{eq:fifteen}
\end{equation}
\begin{equation}
\frac {\partial c}{\partial z} = K-1, \quad z=0,
\label{eq:sixteen}
\end{equation}
\begin{equation}
{\cal S} \frac {\partial \eta}{\partial t} = k \frac {\partial \theta^{(S)}}
{\partial z} - \frac {\partial \theta^{(L)}}{\partial z},
\label{eq:seventeen}
\end{equation}
where the small slope approximation has been invoked in the 
expression for the curvature, $ \kappa \approx - \Delta_r \eta$,
where $\eta$ is the perturbation of the planar position, and
\begin{equation}
F(r) = \frac {\partial U}{\partial z}(r,0) =
2 G_L ({1-\alpha \over 2+\alpha}) 
\sum_{n=1}^{\infty} \frac {r^2-2(nH_0)^2}{[r^2+(nH_0)^2]^{5/2}}.
\label{eq:eighteen}
\end{equation}
The far field conditions for
the perturbations are zero concentration and zero thermal gradients in the
liquid and solid phases. The stability problem is solved using the finite
Hankel transform over  the range $[0,\ell]$ \cite{Pou00}. On using the
approximation $F(r) \sim F(0)$ as $r \rightarrow 0$, we find
\begin{widetext}
\begin{equation}
{\hat \theta}^{(L)} = \Bigl[-[G_L + \sigma \omega^2 + {2 \zeta(3) (\alpha-1)
\over (2+\alpha) H_0^3} G_L] {\hat \eta} + M {\ell (1-K) \over \omega^2}
J_1(\omega \ell) - \lambda^{(L)} L(\omega) \Bigr] e^{- \omega z} 
+ \lambda^{(L)} L(\omega),
\label{eq:nineteen}
\end{equation}
\begin{equation}
{\hat \theta}^{(S)} = \Bigl[ -[ G_S + \sigma \omega^2 + {2 \zeta(3) (\alpha-1)
\over k(2+\alpha)H_0^3} G_L] {\hat \eta} + M {\ell (1-K) \over \omega^2}
J_1(\omega \ell) - \lambda^{(S)} L(\omega) \Bigr]e^{\omega z}
 + \lambda^{(S)} L(\omega),
\label{eq:twenty}
\end{equation}
\begin{equation}
L(\omega) = {\ell G_L J_1(\omega \ell) \over \omega^3} + 
{ 1 \over \omega^2} \int_{0}^{\ell} F(r) r J_0(\omega r)\,dr,
\label{eq:twentyone}
\end{equation}
 \end{widetext}

\begin{figure}
\includegraphics[height=1.8in]{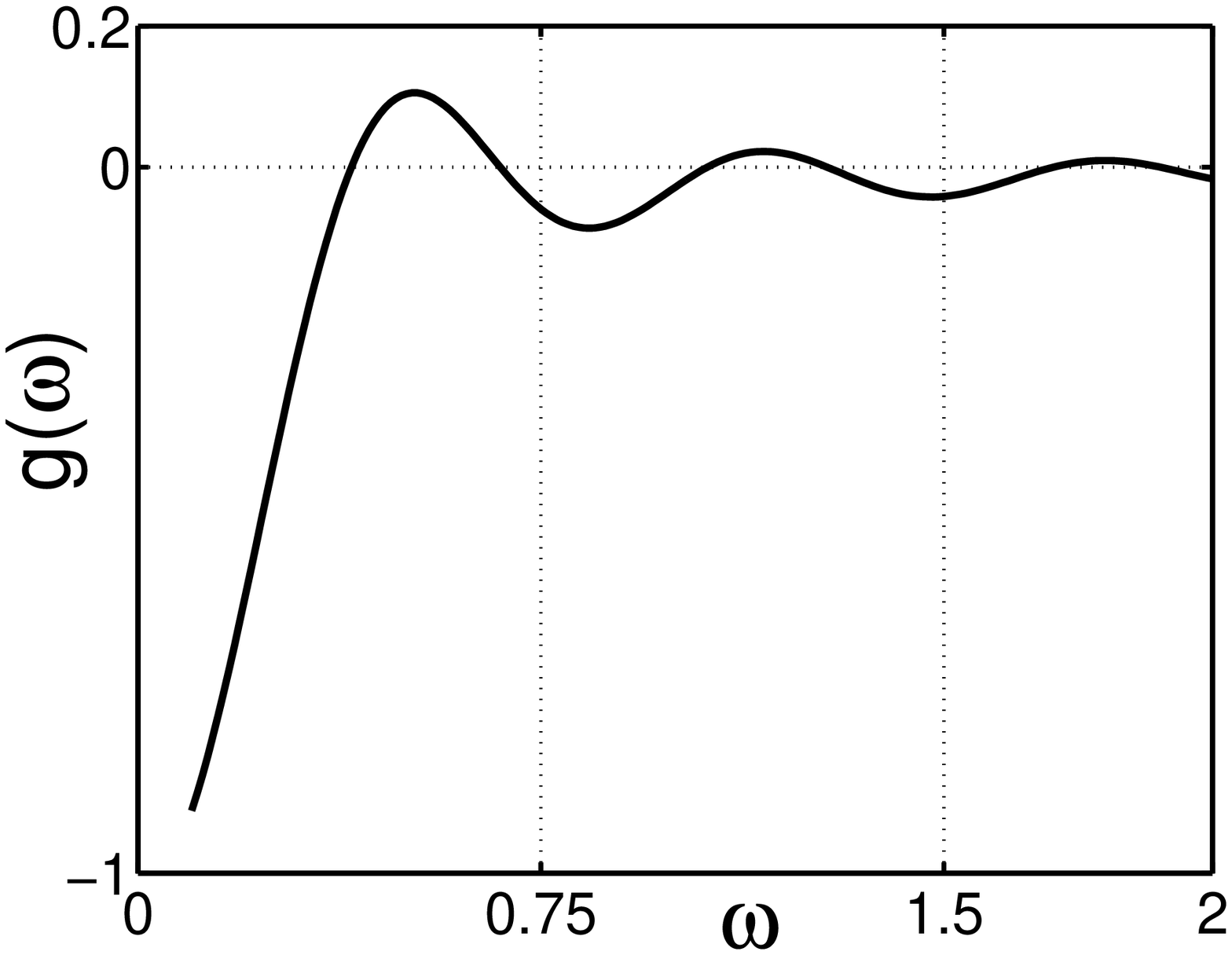}\\
\includegraphics[height=1.8in]{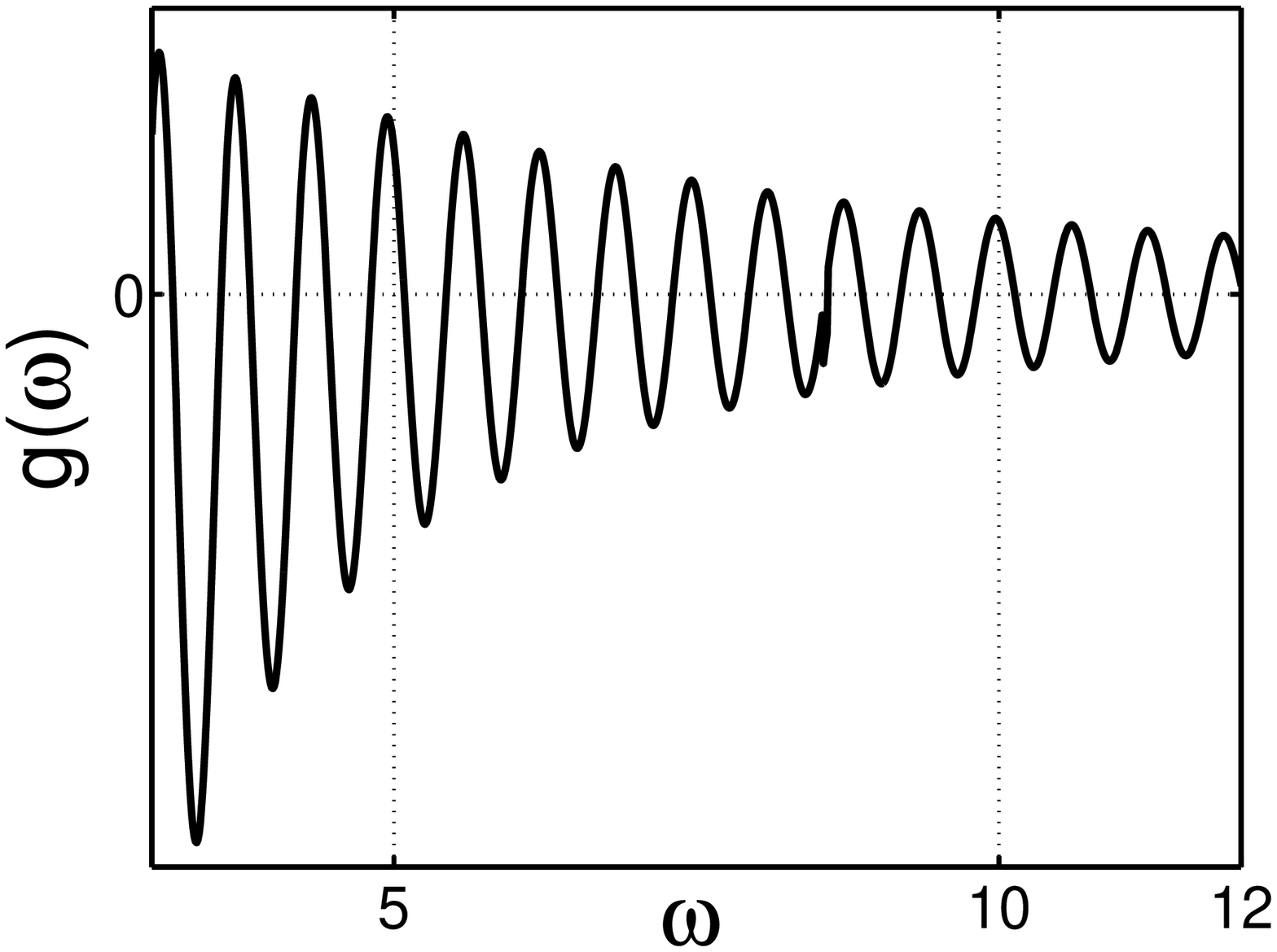}
\caption{\label{fig:epstart}
The small wavenumber behavior (left) and the large wavenumber behavior
(right) of the function $g(\omega)$ for a system consisting of
SCN-ACE containing silicon-carbide (SiC) particles.
The numerical values of the physical constants are \cite{Kur92,Ste95}:
liquidus slope $m=-2.8\rm\,K/wt\%$, segregation coefficient $K=1$,
$C_{\infty}=1.3\rm\,Wt\%$, melting point $T_m=328\rm\,K$, coefficients of
thermal conductivity $k^{(P)} = 85\rm\,W m^{-1} K^{-1}$, $k^{(S)} =
0.225\rm\, W m^{-1} K^{-1}$ and $k^{(L)} = 0.223\rm\, W m^{-1} K^{-1}$
for the particle, solid and liquid, respectively; thermal gradients
$G^{(S)} = 8000\rm\,K m^{-1}$ and $G^{(L)} = 10800\rm\,K m^{-1}$ in the
solid and liquid, repsectively; $\sigma_{sl} = 0.009\rm\,J m^{-2}$;
 $H_0$ and $\ell$ have been arbitrarily chosen
to equal $1.1$ and $10$, respectively; $L = 4.6 \times 10^{7}\rm\,J m^{-3}$.} 
\label{f.2}
\end{figure}

where the hat symbol indicates transformed variables, $\omega$ is the 
wavenumber, $J_m$ is the Bessel function of the first kind of order $m$  and
$\zeta$ stands for the zeta function, i.e. $\zeta(3) = \sum_{n=1}^{\infty}
n^{-3}$.

On imposing the heat balance equation at the interface, Eq. $(18)$,
 an evolution equation 
for the growth of the perturbation ${\hat \eta}$ is obtained. Its solution
yields, 
\begin{equation}
{\hat \eta}(t) = {g(\omega) \over \Lambda(\omega)}+
                 [ {\hat \eta}(0) - {g(\omega) \over \Lambda(\omega)}]
                 e^{-\Lambda(\omega)t /S},
\label{eq:twentytwo}
\end{equation}
where
\begin{equation}
\Lambda(\omega) = {\omega \over S} \Bigl[ {\cal G} + (1+k) \sigma \omega^2
+ {4 \zeta(3) (\alpha-1) G_L \over (2+\alpha)H_0^3} \Bigr],
\label{eq:twentytwob}
\end{equation}
\begin{equation}
g(\omega) = M(1+k) \ell (1-K) {J_1(\omega \ell) \over \omega S}
-{\lambda \omega L(\omega) \over S},
\end{equation}
 ${\cal G} = k G_S + G_L$ is the conductivity weighted thermal gradient,
and $\lambda = \lambda^{(S)} + \lambda^{(L)}$.
The conditions for marginal stability are determined by  setting the
growth rate to zero, i.e. $\Lambda(\omega) =0$. However, when $\Lambda =0$, the
term $g(\omega)/\Lambda(\omega)$ in Eq. $(23)$ is undefined. Therefore,
no meaning to the dispersion relation, $\Lambda (\omega)=0$, could exsit
if $\omega$ did not satisfy the auxiliary equation $g(\omega)=0$.
The ratio, $g\prime(\omega)/\Lambda\prime(\omega)$, where $\prime = d/d\omega$
is however finite.
The function $g(\omega)$ is singular at $\omega=0$ and has infinitely many
zeros $\omega_j$, $j=1,2,3, \ldots$, on the postive $\omega-$axis (Fig. 3).
\begin{figure}
\includegraphics[height=1.8in]{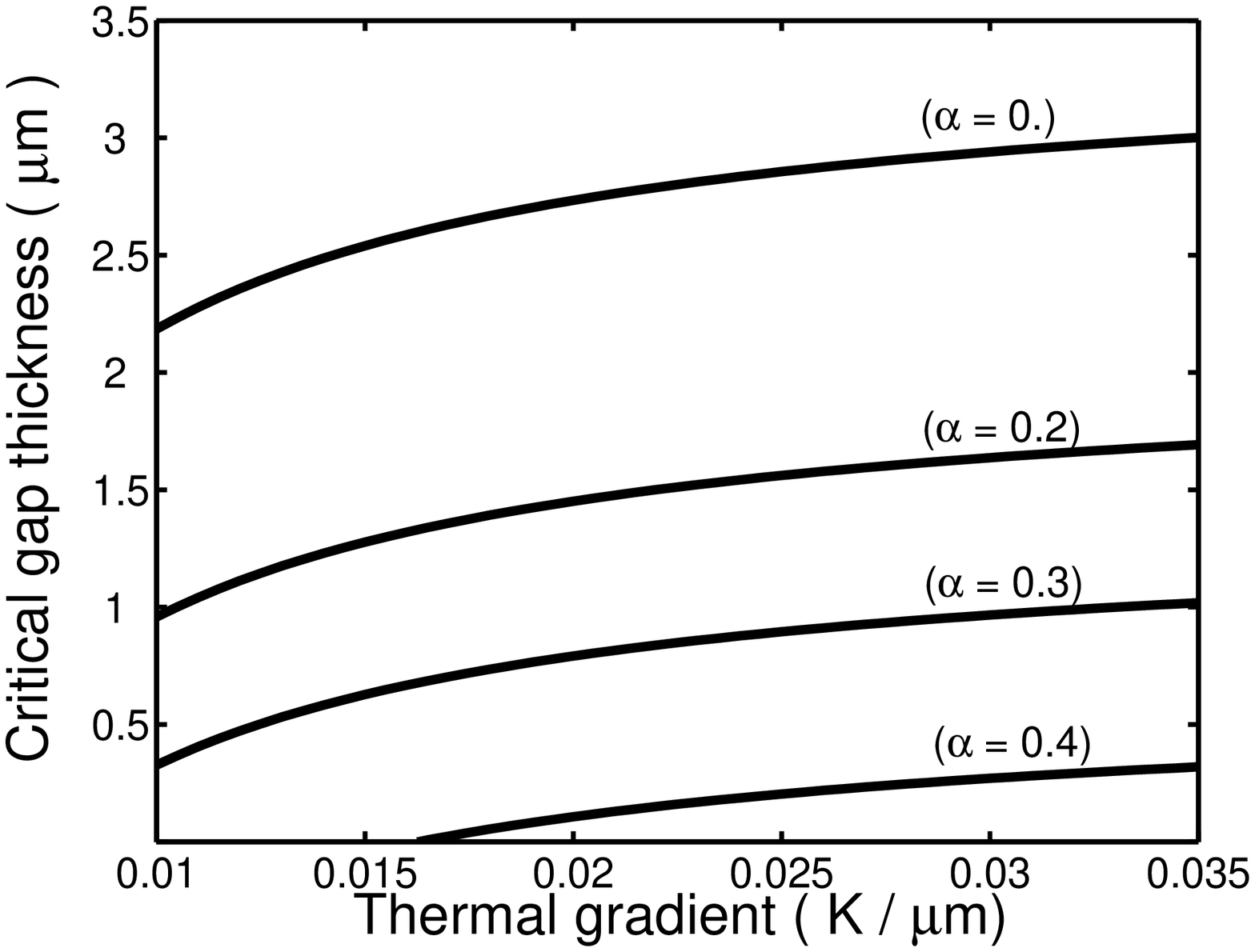}\\
\includegraphics[height=1.8in]{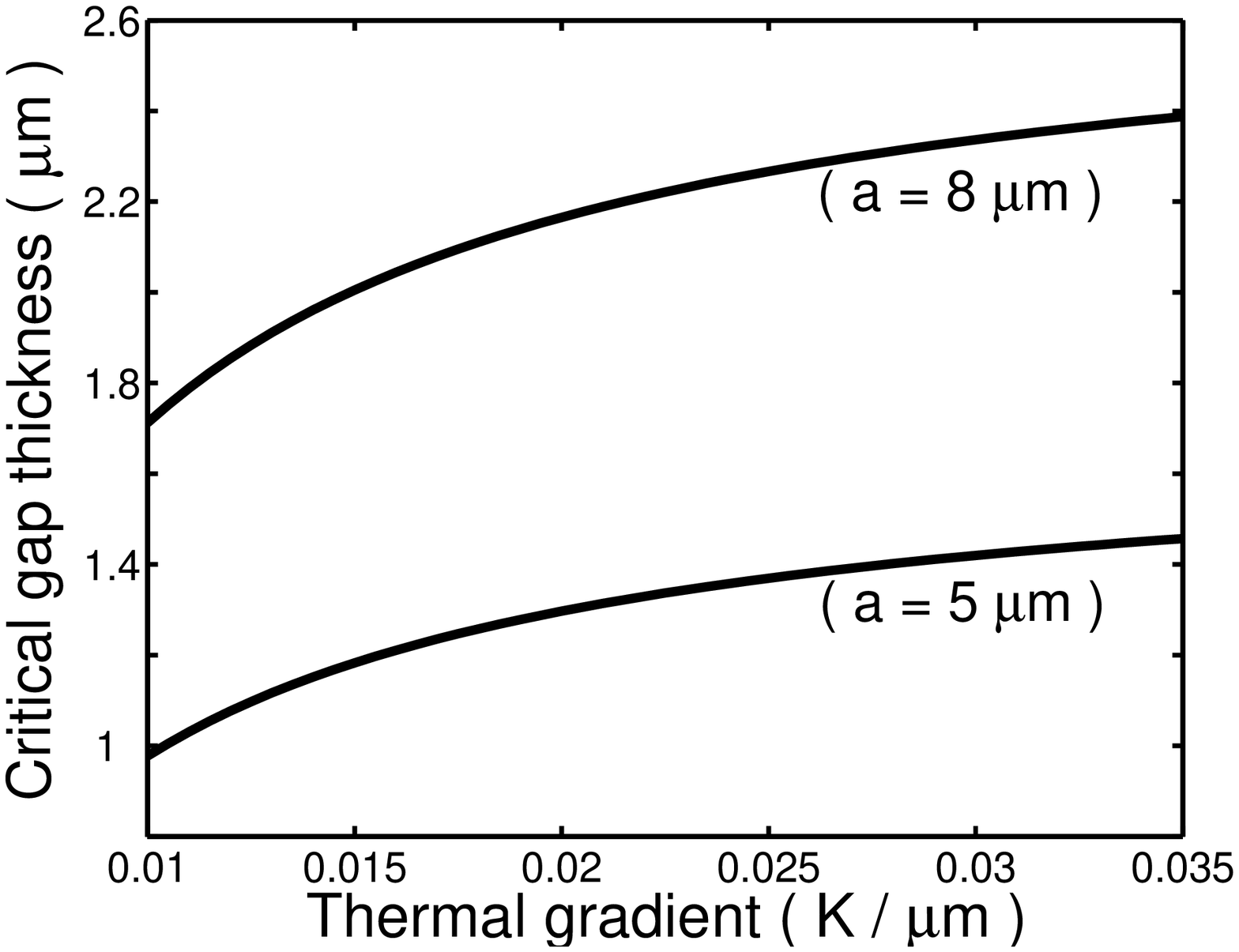}
\caption{\label{fig:epstart}
(left) Plot of the dimensional critical gap thickness, $(d_c -1 )a$,
as function of the dimensional thermal gradient in the melt, $G^{(L)}$, in
unit of Kelvin per micrometer for selected values of the thermal conductance
ratio $\alpha$, (right) plot of the critical gap thickness for 2 insulating
particles of radii $5 \mu m$ and $10 \mu m$ as function of the imposed thermal
 gradient in the melt $G^{(L)}$.}
\label{f.3}
\end{figure}
We set $\omega = \omega_j$ in the dispersion relation $\Lambda(\omega)=0$ and
solve for $H_0$. The critical value of the particle-interface separation,
$d_c$, is the largest value of $H_0$ and is obtained by setting $\omega_j$
equal to the smallest root $\omega_1$. We find
\begin{equation}
d_c = \Bigl[ {4 \zeta(3) (1-\alpha) G_L \over (2+\alpha)[{\cal G} +(1+k)
             \sigma \omega_1^2]} \Bigr]^{1/3}.
\label{eq:twentythree}
\end{equation}
Equation (26) describes the instability criterion in terms of dimensionless
terms. It implies that for prescribed thermal gradients in the liquid and
solid, and for a given thermal conductance ratio $\alpha$, instability of the
planar interface sets in whenever the dimensionless distance from the 
particle's center to the interface, measured by $H_0$, equals $d_c$. 
From the physical standpoint, the instability is manifested only for
parameter values for which $d_c$ exceeds unity and the expression
$(1-\alpha)G_L$ is positive. Figure 4 depicts the range of parameters for
which the instability is observable for a system consisting of SCN-ACE that
is immersed with particles of different sizes and thermal conductivities.
The plot of the dimensional critical gap thickness, $h_{\infty} = (d_c-1)a$,
as a function of the thermal gradient in the melt, $G^{(L)}$, for particles
of different thermal conductivities show that the critical gap thickness
(i) increases with $G^{(L)}$, (ii) decreases with $\alpha$, and (iii) particles
characterized by high values of $\alpha$ require high thermal gradients for the
instability to set in. For increasing particle's size, the critical gap
thickness increases as $\alpha$ and $G^{(L)}$ remain constant.
Furthermore, we note that the
onset of the instability, being of thermal origin, does not depend explicitely
on the concentration. However, the magnitude of $d_c$ is affected by the 
morphological number due to the dependence of the auxiliary function
$g(\omega)$, and thus of the root $\omega_1$, on $M$. Depending on the
physical  process under consideration, the presence or absence of this 
instability can be imposed by the selection of an appropriate set of
physical parameters as dictated by Eq. (26) and shown in Fig. 4.
The instability put forth in this paper also demonstrates the influence a tiny
impurity in the melt can have on the morphological stability of the interface.\\ 
Ahuja \cite{Ahu92} has carried out a detailed experimental study of the
influence of foreign particles on the morphology of a slowly growing
solid-liquid interface under normal gravity conditions 
(see also \cite{Sek91}).
These experiments depict {\it in situ} the real time evolution of the 
solid-liquid interface shape profile in the vicinity of a single particle.
The sequence of photographs depicting the interaction of polystyrene
particles in SCN with an initially planar solid-liquid interface,
shown on page $101$ of Ref. \cite{Ahu92}, provides qualitative confirmation
of the features of the instability put forth in this paper.
The quantitative validation of our prediction requires carefully controlled
experiments in a microgravity environment.


\begin{thebibliography}{99}
\bibitem{Jac58}
K.A. Jackson and B. Chalmers, {\it J. Appl. Phys.} {\bf 29}, 1178 (1958).
\bibitem{Ish94}
H. Ishiguro and B. Rubinsky, {\it Cryobiology} {\bf 31}, 483 (1994).
\bibitem{Gay02}
G. Gay and M.A. Azouni, {\it Crystal Growth $\&$ Design} {\bf 2}, 135 (2002).
\bibitem{End96}
A. Endo, H.S. Chauhan, T. Egi and Y. Shiohara, {\it J. Mater. Sci.} {\bf 11},
795 (2002).
\bibitem{Cly91}
A.R. Kennedy and T.W. Clyne, {\it Cast Metals} {\bf 4}, 160 (1991). 
\bibitem{Uhl64}
D.R. Uhlmann, B. Chalmers and K.A. Jackson, {\it J. Appl. Phys.} {\bf 35},
2986 (1964).
\bibitem{Zub73}
A.M. Zubko, V.G. Lobanov and V.V. Nikonova, {\it Sov. Phys. Crystallgr.}
{\bf 18}, 239 (1973).
\bibitem{Aac77}
A.A. Chernov, D.E. Temkin and A.M. Mel'kinova, 
{\it Sov. Phys. Crystallogr.} {\bf 22} 656-658 (1977).
\bibitem{Sen97}
S. Sen, W.F. Faukler, P. Curreri and D.M. Stefanescu,
{\it Met. Trans. A} {\bf 28}, 2129 (1997).
\bibitem{Had01}
L. Hadji, {\it Phys. Rev. E} {\bf 64}, 051502 (2001).
\bibitem{not01}
The directional solidification of an alloy is subject to a morphological
Mullins-Sekerka instability when the growth rate $V$ exceeds some critical
value (W.W. Mullins and R.F. Sekerka, [{\it J. Appl. Phys.  } {\bf 34}, 323
(1963)]). The planar interface undergoes a bifurcation to a cellular state
whose wavelength is determined from a competition between the destabilizing
diffusion length scale and the stabilizing capillary length scale. The
wavelength varies roughly between $10$ and $100\rm\,\mu m$. 
(I. Durand, K. Kassner, C. Misbah and H. M\"{u}ller-Krumbhaar, [{\it Phys. Rev.
Lett.} {\bf 76}, 3013 (1996)]). This wavelength is of the same order of
magnitude as the size of the inclusions that are used in PMMC and found in 
other processes. In this paper, the mere presence of the particle in the
melt provides the length scale.  Furthermore, the growth rates considered here
are so low ($\epsilon \ll 1$) that the Mullins-Sekerka instability does not
arise.
\bibitem{not02}
Typically, the size of the inclusions vary between $5\rm\, \mu m$ and
$100\rm\, \mu m$, and $D \approx 10^{-9}\rm\, m^2 s^{-1}$. We have considered
here very low growth rates so that the onset of the Mullins-Sekerka instability
is remote  and the particle remains pushed by the interface.
Thus, for $V \approx 10^{-9}\rm\,m s^{-1}$, we have
$\epsilon \approx 10^{-5}$. We assume that the perturbation to the interface
profile is of order ($\epsilon$). The case of a pure substance with zero
growth rate has been analyzed by L. Hadji, [{\it Scripta Materialia} {\bf 48},
665 (2003)].
\bibitem{Car59}
H.S. Carslaw and J.C. Jaeger, {Conduction of Heat in Solids}, second edition
(Clarendon Press, Oxford, 1959).
\bibitem{Che98}
H. Cheng and L. Greengard, {\it SIAM J. Appl. Math,}, {\bf 58}, pp. 122-141
(1998).
\bibitem{Kur92}
W. Kurz and D.J. Fisher, { Fundamentals of Solidification}, third edition
(Trans. Tech. Publications, Switzerland, 1992).
\bibitem{Ste95}  
D. M. Stefanescu, R.V. Phalnikar, H. Pang, S. Ahuja, and B. K. Dhindaw,
{\it SIJ Int.} {\bf 35}, 300 (1995).
\bibitem{Pou00}
A.D. Poularikas, { The Transforms and Applications Handbook}, second edition
(CRC Press, Boca Raton, Florida, 1999).
The interaction of a single particle with the interface sets up a perturbation
in the interface due to the difference in thermal conductivities
$(\alpha \ne 1)$. In case the instability sets in, this disturbance travels
radially in all directions, decreasing in magnitude as it goes.
Far away from the particle, at $r = \ell$, the particle's effect vanishes.
 In our calculations, we have  arbitrarily chosen  $\ell =10$. 
\bibitem{Ahu92}
S. Ahuja, Ph.D. Dissertation, The University of Alabama, p. 100 (1992).
\bibitem{Sek91} 
J.A. Sekhar and R. Trivedi, {\it Mater. Sci. Eng. A} {\bf 147}, pp. 9-21 
(1991). This paper investigates the morphological stability of a
solidifying interface in the presence of large number of particles.
\end{thebibliography}
\end{document}